\documentclass[preprints,article,accept,moreauthors,pdftex]{Definitions/mdpi}

\firstpage{1} 
\pubvolume{xx}
\issuenum{1}
\articlenumber{5}
\pubyear{2019}
\copyrightyear{2019}
\history{Received: date; Accepted: date; Published: date}

\usepackage{amsmath}	
\usepackage{amssymb}	
\usepackage{numprint}
\usepackage[squaren,Gray]{SIunits}
\usepackage{grffile}

\usepackage[T1]{fontenc}
\usepackage{ae,aecompl}

\usepackage{graphicx}	

\newcommand{\rlight}{r_{\rm L}}

\newcommand{\Rs}{R_{\rm s}}

\newcommand{\rot}{\mathbf{\nabla} \times}

\Title{Electrodynamics and radiation from rotating neutron star magnetospheres}

\Author{J\'er\^ome P\'etri$^{1}$} 

\AuthorNames{J\'er\^ome P\'etri}

\address{%
$^{1}$ \quad Universit\'e de Strasbourg, CNRS, Observatoire astronomique de Strasbourg, UMR 7550, F-67000 Strasbourg, France. jerome.petri@astro.unistra.fr
}

\abstract{Neutron stars are compact objects rotating at high speed, up to a substantial fraction of the speed of light (up to 20\% for millisecond pulsars) and possessing ultra-strong electromagnetic fields (close to and sometimes above the quantum critical field of \numprint{4.4e9}~\SIunits{\tesla}). Moreover, due to copious $e^\pm$ pair creation within the magnetosphere, the relativistic plasma surrounding the star is forced into corotation up to the light cylinder where the corotation speed reaches the speed of light. The neutron star electromagnetic activity is powered by its rotation which becomes relativistic in the neighbourhood of this light cylinder. These objects naturally induce relativistic rotation on macroscopic scales about several thousands of kilometers, a crucial ingredient to trigger the central engine as observed on Earth. In this paper, we elucidate some of the salient features of this corotating plasma subject to efficient particle acceleration and radiation, emphasizing several problems and limitations concerning current theories of neutron star magnetospheres. Relativistic rotation in these systems is indirectly probed by the radiation produced within the magnetosphere. Depending on the underlying assumptions about particle motion and radiation mechanisms, different signatures on their light-curves, spectra, pulse profiles and polarisation angles are expected in their broadband electromagnetic emission. We show that these measurements put stringent constraints on the way to describe particle electrodynamics in a rotating neutron star magnetosphere.
}


\begin{document}


\section{Introduction}

Neutron stars are compact objects produced by the explosion of a massive star or by the collapse of an accreting white dwarf reaching the Chandrasekhar limit of about $1.44\,M_\odot$ \cite{chandrasekhar_maximum_1931, landau_theory_1932} where $M_\odot$ is the solar mass. They represent the ultimate fate of the stellar evolution of massive stars before the black hole stage. During their birth, their angular speed and magnetic field are amplified by several orders of magnitude. It is not yet clear what mechanisms are able to produce the expected fields in the range of $B \approx \numprint{e8}-\numprint{e11}~\SIunits{\tesla}$, but a dynamo effect and magnetic flux freezing during the collapse are certainly key processes. Neutron star rotation periods span four decades from several milliseconds to tenths of seconds. The stellar magnetic field drags charged particles into corotation with the star. Relativistic corotating speeds are reached at the light cylinder defined by
\begin{equation}\label{eq:RL}
 \rlight = \frac{c}{\omega} = \frac{c\,P}{2\,\pi} = 48~\SIunits{\kilo\meter} \left( \frac{P}{1~\SIunits{\milli\second}} \right)
\end{equation}
where $c$ is the speed of light, $P=2\pi/\omega$ the pulsar period and $\omega$ its rotation rate.
Moreover, their compactness places them closest to the black hole stage because
\begin{equation}\label{eq:Compacite}
\frac{\Rs}{R} = 0.345 \, \left( \frac{M}{1.4~M_\odot} \right) \, \left( \frac{R}{12~\SIunits{\kilo\meter}} \right)^{-1}
\end{equation}
where $\Rs=2\,G\,M/c^2$ is the Schwarzschild radius, $M$ and $R$ the neutron star mass and radius respectively and $G$ the gravitational constant. Neutron stars are therefore places in the universe where general relativity and quantum electrodynamics act together to sustain their electromagnetic activity.

Simple but realistic orders of magnitude for neutron star rotation periods and magnetic field strengths are easily derived from conservation of angular momentum and magnetic flux during the collapse of the progenitor. These conservation laws imply
\begin{subequations}
	\begin{align}
		M \, \omega \, R^2 & = M_* \, \omega_* \, R_*^2 \\
		\label{eq:flux_conservation}
		B \, R^2 & = B_* \, R_*^2
	\end{align}
\end{subequations}
where the star subscript~$*$ refers to quantities relative to the progenitor. Magnetic field and rotation frequency therefore increase by a factor as large as $R_*^2/R^2 \approx 10^{10}$ if dissipation processes are neglected and all angular momentum and magnetic flux of the progenitor go to the neutron star. The magnetic flux conservation argument, equation (\ref{eq:flux_conservation}), was first discussed by \cite{woltjer_x-rays_1964}. The above estimates are however probably largely overestimated because not all of the progenitor is collapsing and because its mass is not conserved during the implosion \cite{spruit_origin_2008}. Only the iron core of a massive star produces a neutron star star whereas the outer shells are expelled. Electrodynamics in a relativistically rotating frame is a crucial ingredient in our understanding and modelling of neutron star magnetospheres. In this paper, we discuss some issues about rotating magnetospheres and their radiative properties.

First we briefly remind the rotating coordinate system used and its implication for field transformations and especially for Maxwell equations in section~\ref{sec:RotatingFrame}. Next we discuss the electromagnetic field expected inside and outside a neutron star in section~\ref{sec:ElectromagneticFiled}. The motion of charged particles in this field is exposed in section~\ref{sec:ParticleMotion}. Such trajectories can be computed in solutions found from numerical simulations as explained in section~\ref{sec:NumericalSimulations}. However, corotation is not compulsory and differentially rotating magetospheres have been found as shown in section~\ref{sec:DifferentialMagnetospheres}. Some clues about the electrodynamics of pulsar magnetospheres can be gained from their radiation as explained in section~\ref{sec:Radiation}. Possible extensions to general relativity and multipolar magnetic fields are discussed in Sec.~\ref{sec:Discussion}. We conclude with a brief summary in section~\ref{sec:Conclusions}.

 

\section{Rotating vs inertial frame}
\label{sec:RotatingFrame}

Neutron stars are mainly observed through their pulsed emission detected in the radio wavelength \cite{manchester_observed_1981} but also at very high-energy by gamma-ray photons in the MeV/GeV band \cite{abdo_second_2013} or through thermal X-ray emission from hot spots on the surface \cite{guillot_nicer_2019}. Such emission is attributed to ultra-relativistic charged particles flowing inside the magnetosphere. The very stable and periodic pulsation is explained by the stellar rotation. It is believed that these particles corotate with the star almost up to the light cylinder and generate a relativistic magnetized outflow outside the light cylinder known as the pulsar wind \cite{goldreich_pulsar_1969}. Description of the flow taking properly into account this corotation is therefore central to our understanding of neutron star magnetospheric emission. Before going into the dynamics of this plasma and its underlying particle trajectories let us briefly review relativistic rotating frames and related electrodynamics problems. An excellent review about relativistically rotating frames can be found in \cite{rizzi_relativity_2003}.

\subsection{Metric}

The space-time geometry of a rotating system is best described in a cylindrical coordinate system labelled by  $(t,r,\varphi,z)$. The coordinate transformation from an inertial observer~$(t,r,\varphi,z)$ to an observer rotating at the stellar angular speed~$(t',r',\varphi',z')$ is usually written as
\begin{equation}
\label{eq:TransformCoordonnees}
t'=t \qquad ; \qquad r'=r \qquad ; \qquad \varphi' = \varphi - \omega\,t \qquad ; \qquad z'=z .
\end{equation}
It is essential to realise that this transformation does not lead to a new orthonormal basis in the rotating frame. Therefore physical quantities measured locally by an observer in this coordinate system can not be directly read off from these basis vectors. Indeed, the space-time geometry of an uniformly rotating frame is given in the rotating observer frame by the metric
\begin{equation}\label{eq:Metric}
ds^2 = c^2\,dt^2 - dr^2 - \frac{r^2\,d\varphi^2}{1-r^2\,\omega^2/c^2} - dz^2 .
\end{equation}
The coordinates~$(t',r',\varphi',z')$ are not related to any observer because they do not form an orthonormal basis. In order to relate measurements between the inertial and the rotating observers, it is more convenient to introduce two orthonormal bases attached to each observer. The transformation from the inertial to the rotating frame is simply a Lorentz boost as is always the case when performing reference frame transformations between orthonormal bases. Explicitly, the orthonormal basis vectors of the rotating frame are
\begin{subequations}
	\label{eq:BaseTournante}
	\begin{align}
	\mathbf{e}_0' & = \Gamma \left( \mathbf{e}_t + \beta \, ( - \sin\Phi \, \mathbf e_x + \cos\Phi \, \mathbf e_y ) \right) = \Gamma \, ( 1, - \beta \, \sin\Phi, \beta \, \cos\Phi, 0 ) \\
	\mathbf{e}_1' & = \cos \Phi \, \mathbf e_x + \sin \Phi \, \mathbf e_y = (0,  \cos \Phi, \sin \Phi, 0) \\
	\mathbf{e}_2' & = \Gamma \, \left( \beta \, \mathbf{e}_t - \sin\Phi \, \mathbf e_x + \cos\Phi \, \mathbf e_y \right) = \Gamma \,( \beta , - \sin\Phi , \cos\Phi, 0 ) \\
	\mathbf{e}_3' & = \mathbf{e}_z = (0,0,0,1)
	\end{align}
\end{subequations}
where the phase is $\Phi=\omega\,t$, the relative speed is $\beta = \frac{\omega\,r}{c}$, and the Lorentz factor $\Gamma = (1-\beta^2)^{-1/2}$. 
This coordinate transformation is of Lorentz type, going from a Minkowskian metric to another Minkowskian metric. Building on this orthonormal basis, it is easy to deduce the relation between the electromagnetic fields measured by the two observers as we now show.

\subsection{Electrodynamics}

The electromagnetic field tensor in an inertial frame is derived from the electromagnetic quadri-potential $A_i = (\phi/c, - \mathbf{A})$ ($\phi$ being the scalar potential and $\mathbf{A}$ the vector potential adopting the metric with signature (+,-,-,-)) by $F_{ik} = \partial_i A_k - \partial_k A_i$. The electromagnetic field tensors in the inertial and rotating frame are related by a special relativistic transformation according to the previous section from the basis transformation eq.~(\ref{eq:BaseTournante}).
They are also found from the definition using the observer 4-velocity~$\mathbf{u}$ by projection of the electromagnetic field tensor~$F_{ik}$ and its dual ${^*\!F}_{ik} = \frac{1}{2} \, \epsilon_{ikmn} \, F^{mn}$ \cite{jackson_electrodynamique_2001} onto the observer world line such that
\begin{subequations}
\begin{align}
	E_i/c & = F_{ik} \, u^k \\
	B_i & = {^*\!F}_{ik} \, u^k .
\end{align}
\end{subequations}
Explicit computations of these transformations show that it is simply the Lorentz transformation of the electromagnetic field between two inertial observers with relative velocity $\mathbf{V} = r \, \omega \, \mathbf{e}_\varphi$. Introducing the normalized velocity by $\boldsymbol{\beta} = \mathbf{V}/c$, the transformations of the electric and magnetic field vectors are
\begin{subequations}
	\begin{align}
	\mathbf{E}' & = \Gamma \, \left[ \mathbf{E} - \frac{\Gamma}{\Gamma+1} \, ( 
	\boldsymbol{\beta} \cdot \mathbf{E}) \, \boldsymbol{\beta} + c \, \boldsymbol{\beta} \wedge 
	\mathbf{B} \right] \\
	\mathbf{B}' & = \Gamma \, \left[ \mathbf{B} - \frac{\Gamma}{\Gamma+1} \, ( 
	\boldsymbol{\beta} \cdot \mathbf{B}) \, \boldsymbol{\beta} - \boldsymbol{\beta} \wedge \mathbf{E}/c 
	\right]
	\end{align}
\end{subequations}
where primed quantities are defined in the rotating frame. There is nothing special about rotation if transformations are made locally between inertial observers. This holds for aberration and Doppler effects when emission emanates from within the light cylinder ($r<\rlight$). These results are easily extended to general relativity when gravitation is included. However special relativity applies if transformations are made \textit{locally} between inertial observers \cite{gourgoulhon_relativite_2010}.

\subsection{Doppler effect}

The Doppler effect is subject to the same transformation as the electromagnetic field. A simple Lorentz transformation between inertial frames holds to relate photon propagation direction~$\mathbf{n}=(n_r,n_\theta,n_\varphi)$ and frequency~$\nu$ in the observer frame and in the instantaneous inertial frame coinciding locally with the corotating frame. It can be checked by a direct derivation from the transformation of coordinates between both frames as given by eq.~(\ref{eq:BaseTournante}). For the sake of completeness, they are given in a spherical coordinate systems~$(r,\vartheta,\varphi)$ by
\begin{subequations}
	\begin{align}
	\nu & = \gamma_{\rm obs} \, \nu' \, ( 1 + \beta_{\rm obs} \, n_\varphi' ) \\
	n_r & = \frac{n_r'}{\gamma_{\rm obs} \, ( 1 + \beta_{\rm obs} \, n_\varphi' )} \\
	n_\vartheta & = \frac{n_\vartheta'}{\gamma_{\rm obs} \, ( 1 + \beta_{\rm obs} \, n_\varphi' )} \\
	n_\varphi & = \frac{\beta_{\rm obs} + n_\varphi'}{1 + \beta_{\rm obs} \, n_\varphi'} .
	\end{align}
\end{subequations}
This aberration formula also holds in general relativity if the Lorentz factor $\gamma_{\rm obs}$ and velocity $\beta_{\rm obs}$ are properly defined as the true physical quantities measured by a corotating observer. Mathematically, this requires to switch from a general curvilinear coordinate system, like the metric of a rotating disk, to an orthonormal coordinate system associated to the Minkowskian metric. Only the latter coordinates have a clear physical interpretation, the former being only appropriate (or not) coordinates to describe the problem.

To summarize, in any local Lorentzian frame, using an orthonormal basis, the Doppler factor is written geometrically as
\begin{equation}\label{eq:Doppler}
\mathcal{D} = \frac{1}{\Gamma \, (1 - \boldsymbol\beta \cdot \mathbf{n})} .
\end{equation}
It enables to relate photon propagation directions in both frames as
\begin{equation}
\label{eq:AberrationLorentz}
\mathbf{n} = \frac{1}{\mathcal{D}} \, \left[ \mathbf{n}' + \Gamma \, \left( \frac{\Gamma}{\Gamma+1} \, ( \boldsymbol\beta \cdot \mathbf{n}' ) + 1 \right) \, \boldsymbol\beta \right]
\end{equation}
with the usual redshift phenomenon relating the frequency in the rotating frame~$\nu'$ to the frequency in the observer frame~$\nu$ by
\begin{equation}\label{eq:DopplerOmega}
\nu = \mathcal{D} \, \nu' .
\end{equation}
The aberration effect can only be computed for a physical realisation of a corotating frame. It would fail right at the light cylinder or outside it. Unfortunately, a smooth transition from the magnetosphere $r\leq\rlight$ to the wind $r\geq\rlight$ is required for modelling the pulsar broad band emission. This tells us that the introduction of a corotating frame where the electromagnetic field and plasma motion are both stationary will not help in advancing our understanding of neutron star electrodynamics. It is preferable to keep the physical quantities expressed in an inertial frame even if the coordinate system can be advantageously described in a corotating frame (not related to any comoving observer). We expose such a technique in the next section for the evolution of the electromagnetic field.

\subsection{Maxwell equations in a rotating coordinate system}

Measuring the electromagnetic field in the corotating system is possible up to the light-cylinder. However, outside this surface, the metric has no physical significance any more. It is impossible to describe the neutron star electrodynamics in whole space with the field locally measured by a corotating observer because such an observer does not exist when $r\geq\rlight$.

Nevertheless, it is relevant and useful to keep the definition of the electromagnetic field as measured in the inertial reference frame but using the rotating coordinate system to localize it with~$(t',r',\varphi',z')$. In such a case the time derivative of any vector field~$\mathbf{A}$ is given by \cite{beskin_mhd_2009}
\begin{equation}
\label{eq:Derivee}
\frac{\partial \mathbf A}{\partial t} = \frac{\partial \mathbf A}{\partial t'} + \textrm{curl} \, ( \mathbf{V}_{\rm rot} \wedge \mathbf A ) - \mathbf{V}_{\rm rot} \, \textrm{div} \mathbf A .
\end{equation}
The solid body corotation velocity, expressed in the inertial frame, is simply
\begin{equation}
\label{eq:VitesseCorotation}
\mathbf{V}_{\rm rot} = \boldsymbol{\omega} \wedge \mathbf{r} = r\,\omega\, \mathbf{e}_\varphi .
\end{equation}
Note that it is not restricted to remain less than the speed of light. There is no singularity in eq.~(\ref{eq:Derivee}) when crossing the light cylinder. With the correspondence established in eq.~(\ref{eq:Derivee}), in the rotating coordinate system, Maxwell equations become
\begin{subequations}
	\label{eq:MaxwellCorotation}
	\begin{align}
	\frac{\partial \mathbf{B}}{\partial t'} & = - \, \textrm{curl} \, (  \mathbf{E} + \mathbf{V}_{\rm rot} \wedge \mathbf B ) \\
	\frac{\partial \mathbf{E}}{\partial t'} & = \textrm{curl} \, ( c^2 \, \mathbf{B} - \mathbf{V}_{\rm rot} \wedge \mathbf E ) - \frac{\mathbf{j}}{\varepsilon_0} + \mathbf{V}_{\rm rot} \, \textrm{div} \mathbf E .
	\end{align}
\end{subequations}
Note however the subtleties that $\mathbf{E}$ and $\mathbf{B}$ are still defined as observed in the inertial frame. No particular problem arises at the light cylinder when Maxwell equations are written in this way. In section~\ref{sec:NumericalSimulations} about numerical simulations, we use this mathematical formulation to solve for the force-free (FFE) and radiative magnetosphere for an oblique rotator as a typical example. But first we have to define the electromagnetic field inside and outside the star in vacuum.

\section{Electromagnetic field inside and outside the star}
\label{sec:ElectromagneticFiled}

To first approximation, a neutron star can be assimilated to a very good conductor. We therefore assume that the electric field inside the star as seen by an observer at rest with respect to the star, the so called comoving observer, vanishes. In the inertial frame of a distant observer, the electric field~$\mathbf{E}$ is given by the usual Lorentz transformation
\begin{equation}
\label{eq:Einterieur}
\mathbf{E} + \mathbf{V}_{\rm rot} \wedge \mathbf{B} = \mathbf{0}
\end{equation}
where $\mathbf{B}$ is the magnetic field in the same observer frame. At this stage, already several implicit assumptions must be done. Should we assume a prescribed magnetic field in the observer frame~$\mathbf{B}$ or in the corotating frame~$\mathbf{B}'$? Both situations are obviously not identical. If the magnetic field is fixed in the observer frame, then in the star corotating frame we find
\begin{equation}
\label{eq:TransformB}
\mathbf{B}' = \frac{\mathbf{B}}{\Gamma_{\rm rot}} + \frac{(\boldsymbol\beta_{\rm rot} \cdot \mathbf{B})}{\Gamma_{\rm rot}+1} \, \boldsymbol\beta_{\rm rot}
\end{equation}
with $\Gamma_{\rm rot} = (1-\boldsymbol\beta_{\rm rot}^2)^{-1/2}$. If the magnetic field is fixed in the rest frame of the star, which seems more reasonable, then the observer will measure a magnetic field
\begin{equation}
\label{eq:TransformBp}
\mathbf{B} = \Gamma_{\rm rot} \, \left[ \mathbf{B}' - \frac{\Gamma_{\rm rot}}{\Gamma_{\rm rot}+1} \, (\boldsymbol\beta_{\rm rot} \cdot \mathbf{B}') \, \boldsymbol\beta_{\rm rot} \right] .
\end{equation}
Note that both assumptions leads to $\boldsymbol\beta_{\rm rot} \wedge \mathbf{B} = \Gamma _{\rm rot}\, \boldsymbol\beta_{\rm rot} \wedge \mathbf{B}'$ so eq.~(\ref{eq:Einterieur}) remains valid in any case inside the perfectly conducting star. Practically, the corotation speed inside the star is always weakly or mildly relativistic, therefore $\beta_{\rm rot} \ll1$ reducing both approaches to $\mathbf{B} \approx \Gamma_{\rm rot} \, \mathbf{B}'$. An exact analytical solution for the radiating electromagnetic field has been given by \cite{deutsch_electromagnetic_1955} assuming a dipolar magnetic field in the inertial frame and neglecting relativistic effects. Close to the light-cylinder, relativistic effects have been taken into account as investigated by \cite{belinsky_radiation_1994}. Moreover, the current induced inside the star by its rotation influences the magnetic field itself as shown by \cite{roberts_electromagnetic_1979}.

Consequently, the description of the magnetic field inside the star is already biased by some assumptions on its geometry in the rotating or inertial frame. What happens outside, in its magnetosphere?  Also, at large distances, outside the light cylinder $r>\rlight$ there exist no more physical frame in corotation with the star. The Lorentz transformation of the electromagnetic field does not apply anymore. It is preferable to stay in the observer frame without any reference to a corotating frame.

The strong electric field induced by the rotating star pulls out charged particles, filling the magnetosphere with relativistic electron/positron pairs \cite{goldreich_pulsar_1969}. To good accuracy, this magnetosphere is assumed to be in force-free equilibrium, neglecting particle inertia and fluid temperature, as well as gravity because of the electromagnetic field strength producing forces several orders of magnitude stronger than the gravitational attraction. The magnetosphere is therefore set into corotation with the star up to the light cylinder at a radius $r=\rlight$. Outside this cylinder, plasma corotation cannot be maintained. The flow of leptons generates space charges and currents that significantly modify the electromagnetic structure. The charge density~$\rho_{\rm e}$ produced by the electric field, following Maxwell-Gauss law and the perfect conductor hypothesis eq.~(\ref{eq:Einterieur}) diverges right at the light-cylinder because
\begin{equation}\label{eq:densite_charge}
\rho_{\rm e} = \varepsilon_0 \, \textrm{div} \, \mathbf{E} = - 2\,\varepsilon_0\,\frac{\boldsymbol\omega \cdot \mathbf{B}}{1 - r^2/\rlight^2}
\end{equation}
unless $\boldsymbol\omega \cdot \mathbf{B}=0$ on this surface. The magnetic field must therefore adjust itself in order to keep the constrain $\boldsymbol\omega \cdot \mathbf{B} = 0$ at $r=\rlight$ which implies $B_{\rm z}(\rlight) = 0$. Another possibility would be to break the corotation approximation by introducing some dissipation through an ad-hoc resistivity or through a radiative dissipation mechanism as shown in subsequent sections. Once the electromagnetic field settled down, we need to investigate particle motion in these fields.

\section{Particle motion in the magnetosphere}
\label{sec:ParticleMotion}

In order to predict the radiation emanating from neutron star magnetospheres, a good understanding of particle trajectories in the electromagnetic field produced by these stars is required. Several attempts focused exclusively on motion restricted to within the light-cylinder. This limitation avoids the problem of transformations involving larger than the speed of light relative velocities between inertial frames. Radiation from outside the light cylinder therefore requires another description. This artificial transition between the magnetosphere and the wind is far from satisfactory. Claims have been made about a new technique to compute emission everywhere, for instance in the vacuum Deutsch field \cite{bai_uncertainties_2010}, but we will show that at least in some cases it fails too to smoothly join inside and outside light cylinder electrodynamics. Moreover, their model based on the assumption that eq.(\ref{eq:Einterieur}) is valid within the magnetosphere is inconsistent with the vacuum assumption of a Deutsch field. A more satisfactory solution includes the plasma feedback self-consistently as proposed by \cite{bai_modeling_2010} in order to avoid superluminal particle speed everywhere. In the following subsections, we summarize the most studied maybe not the most effective models of particle trajectories.

Three different approximations for the particle motion have been tried. Indeed a charged particle can be seen as 
\begin{itemize}[leftmargin=*,labelsep=5.8mm]
\item following magnetic field lines~$\mathbf{B'}$ in the corotating frame.
\item following magnetic field lines~$\mathbf{B}$ in the inertial frame in addition to a corotation imposed by the stellar rotation.
\item following the ultra-relativistic radiation reaction limit leading to the so-called Aristotelian dynamics. Acutally, this limit can be explained by Newtonian dynamics in a stationary regime balancing electric acceleration and radiation friction.
\end{itemize}
These different views are not equivalent to each other because they assume different electromagnetic field structures, either fixed in the rotating frame or in the inertial frame. Let us discuss these approaches in depth starting with the corotating frame view.

\subsection{Corotating frame}
\label{subsec:frame}

Viewed from the corotating frame, particles are assumed to follow magnetic field lines along~$\mathbf{B}'$. Therefore in this frame the particle velocity is given by
\begin{equation}
\label{eq:VitesseRefRotation}
\mathbf{v}' = v' \, \frac{\mathbf{B'}}{B'} = v' \, \mathbf{n}'_{\rm B}
\end{equation}
where $v'$ is the particle speed along the field line $\mathbf{B}'$ in the normalized direction~$\mathbf{n}'_{\rm B} = \mathbf{B'}/B'$. How then to choose this field~$\mathbf{B}'$? Some authors used in the past the relation $\mathbf{B}' = \mathbf{B}$ which is only correct to second order in $\beta_{\rm rot}$, a results derived from the coordinate transformation between inertial frame and rotating coordinate systems. However, this equality was used by several authors to compute pulsar high-energy light-curves at high altitude, up to a substantial fraction of the light-cylinder \cite{romani_gamma-ray_1995,  dyks_two-pole_2003, dyks_relativistic_2004, cheng_three-dimensional_2000}. Light curves and sky maps derived from this model are sensitive to the upper boundary of the radiating zone. However this dependence is undesirable. Moreover, the rotating coordinate system is not an orthonormal basis, therefore $\mathbf{B}'$ should not be interpreted as the local magnetic field measured by a rotating observer. It must be computed according to the Lorentz transformation \cite{bai_uncertainties_2010}. Nevertheless, both descriptions agree to good accuracy for non relativistic corotating speeds.

Going into the rotating frame synchronous with the neutron star rotation can lead to misinterpretation of the physical electromagnetic field measured by a local observer. Moreover, the corotating frame can not be extended beyond the light cylinder radius~$\rlight$. Such description therefore faces severe difficulties to deal with the entire neutron star magnetosphere and is inadequate to efficiently model them from the surface to large distances within the striped wind $r\gtrsim\rlight$. Much better we think is to perform all calculations in the inertial frame of a distant observer as we now describe in the next two sections.

\subsection{Corotating velocity}
\label{subsec:velocity}

When staying in the observer frame, without reference to any rotating frame, the velocity is described by a velocity component along the magnetic field lines~$\mathbf{B}$, now expressed in the inertial frame and denoted by $v_\parallel$, and a velocity component due to the dragging by the star denoted by~$\mathbf{V}_{\rm rot}$. Therefore we write
\begin{equation}
\label{eq:VitesseParticule}
\mathbf{v} = v_\parallel \, \frac{\mathbf{B}}{B} + \mathbf{V}_{\rm rot} . 
\end{equation}
The velocity along the field line~$v_\parallel$ must be chosen in order to keep the total speed smaller than the speed of light, $v<c$. It requires a special configuration of the magnetic field~$\mathbf{B}$ with an increasing toroidal component to compensate for the linear increase in corotation speed as given by $\mathbf{V}_{\rm rot}$. Therefore not all magnetic field configurations are permitted to fulfil this constrain.

By assumption, in some models \cite{romani_gamma-ray_1995,  dyks_two-pole_2003, dyks_relativistic_2004, cheng_three-dimensional_2000}, particles follow magnetic field lines in the corotating frame. Their distribution function is isotropic in the rest frame of the fluid. Following the previous prescriptions by \cite{bai_uncertainties_2010}, we assume that their Lorentz factor~$\Gamma$ is constant in the observer frame such that the velocity, being a combination between propagation along field lines and corotation at speed~$\mathbf{V}_{\rm rot}$, is
\begin{equation}
\label{eq:AberrationCorot}
\mathbf{v} = v_\parallel^{\rm c} \, \mathbf{t} + \mathbf{V}_{\rm rot}
\end{equation}
where $\mathbf{t} = \pm \mathbf{B}/B$ is the outward pointing tangent vector to the field line. Solving for the parallel velocity $v_\parallel^{\rm c}$ the only real and positive solution is
\begin{equation}
\label{eq:SolutionVparallel}
v_\parallel^{\rm c} = - \mathbf{t} \cdot \mathbf{V}_{\rm rot} + \sqrt{(\mathbf{t} \cdot \mathbf{V}_{\rm rot})^2 + v^2 - V_{\rm rot}^2} .
\end{equation}
$V_{\rm rot}$ exceeds the speed of light outside the light cylinder by definition. The term $v^2 - V_{\rm rot}^2$ in the square root becomes negative and must be compensated by the term $(\mathbf{t} \cdot \mathbf{V}_{\rm rot})^2$ meaning that the magnetic field must be strongly bend toward the azimuthal direction $\mathbf{e}_\varphi$. The Deutsch field does not satisfy this requirement and cannot be used to study photon emission within the wind if this view is adopted. Knowing the velocity, we get the Doppler factor for radiation as explained in eq.~(\ref{eq:Doppler}). This velocity field assumes that the electric field vanishes in the corotating frame. But this requires a large amount of plasma to screen the electric field, in contradiction with the vacuum assumption made in \cite{bai_uncertainties_2010}. Therefore, the aberration formula eq.~(\ref{eq:AberrationCorot}) can only be an approximation in this case. Moreover, this approximation also fails at sufficiently large distances because the Deutsch field solution~\cite{deutsch_electromagnetic_1955} possesses a magnetic field structure for which the polo\"\i dal component does not decay fast enough with respect to the toroidal component. Real solutions to eq.~(\ref{eq:SolutionVparallel}) do not exists at several light-cylinder radii because the square root in eq.~(\ref{eq:SolutionVparallel}) becomes negative. Indeed, taking an orthogonal rotator, it can be shown that in the equatorial plane the term in the square root of eq.~(\ref{eq:SolutionVparallel}) tends to $v^2-4\,c^2<0$ for $r\rightarrow+\infty$ on the spiral given by $\varphi + r/\rlight - \omega\,t = \pi/2$. In the most favourable case for which $v=c$, it actually becomes negative already at the light-cylinder. Using the corotating frame does not help to go beyond the light-cylinder for vacuum fields. Nevertheless, the description exposed in this section is applicable to force-free magnetospheres that exactly cancel the electric field in the frame comoving with the plasma at the electric drift speed. Only in such FFE models can this prescription be correctly applied in whole space, within the magnetosphere $(r\le\rlight$) and within the wind $(r\ge\rlight$).

Is it possible to find a formulation alleviating the need for special magnetic field configurations? In our opinion, there exist a simple and efficient way to compute particle trajectories in any electromagnetic field when moving at the radiation reaction limit. We detail this last approach in the next section.

\subsection{Aristotelian dynamics}
\label{subsec:aristote}

Particles in the neutron star magnetosphere are ultra-relativistic. They copiously radiate photons during their motion. This has to be taken into account. The simplest approximation is given by the radiation reaction limit, where the radiation force, acting as a damping working against the motion, a kind of radiative friction, exactly compensates for the electric acceleration. It is sometime called Aristotelian electrodynamics because the velocity is completely and solely determined by the electromagnetic field felt locally by the particles although we believe that it is more appropriate to speak about radiation reaction motion because it can be derived from the Lorentz force with radiative friction and this according to Newtonian dynamics. The expression for the velocity has been derived in \cite{herold_generation_1985}, but see also \cite{mestel_stellar_1999}. Assuming that particles move exactly at the speed of light (which is an excellent approximation in neutron star magnetospheres), depending on the sign of their charge, their velocity reads
\begin{equation}
\label{eq:VRR}
\mathbf{v}_\pm = \frac{\mathbf{E} \wedge \mathbf{B} \pm ( E_0 \, \mathbf{E} / c + c \, B_0 \, \mathbf{B})}{E_0^2/c^2+B^2}
\end{equation}
where the plus sign corresponds to positive charges and the minus sign to negative charges. Actually, the velocity is independent of the mass~$m$ over charge~$q$ ratio~$q/m$, it only depends on the sign of its charge. Moreover, we introduced the electromagnetic field strengths $E_0$ and $B_0$ according to the two electromagnetic invariants~$(\mathcal{I}_1,\mathcal{I}_2)$ such that 
\begin{subequations}
	\begin{align}
	\mathcal{I}_1 & = \mathbf{E}^2 - c^2 \, \mathbf{B}^2 = E_0^2 - c^2 \, B_0^2 \\
	\mathcal{I}_2 & = c\,\mathbf{E} \cdot \mathbf{B} = c\,E_0 \, B_0
	\end{align}
\end{subequations}
with the subsidiary condition $E_0 \geqslant 0$ ensuring that the radiation reaction force is always directed oppositely to the velocity direction. As explained in \cite{petri_general-relativistic_2018} these invariants are related to the electromagnetic field strength in a frame where $\mathbf{E}$ and $\mathbf{B}$ are parallel. The lepton motion can be decomposed into an electric drift part along the vector $\mathbf{E} \wedge \mathbf{B}$, a motion along magnetic field lines~$\mathbf{B}$ and a motion along electric field lines~$\mathbf{E}$. This last part of the motion is responsible for dissipation because the power of the Lorentz force is $q\,(\mathbf{E} + \mathbf{v}_\pm \wedge \mathbf{B}) \cdot \mathbf{v}_\pm = q \, \mathbf{v}_\pm \cdot \mathbf{E} = Z\,e\,c\,E_0 \geq 0$ where $q=\pm Z\,e$ depending on the charge~$Z$ of the particle: positrons and electrons have $Z=1$ whereas ions have $Z$ arbitrary. The velocity field (\ref{eq:VRR}) is regular in whole space, nothing singular happens at the light-cylinder. It can be implemented to compute realistic pulsar light-curves and spectra even in vacuum Deutsch solution as shown by \cite{petri_pulsar_2019}. This latest work serves as a starting point to investigate more deeply pulsar magnetospheric radiation by including for instance the plasma feedback onto the Deutsch field as will be shown in section~\ref{sec:NumericalSimulations}.

In the near field zone, i.e. close to the neutron star surface, where $E\ll c\,B$, the particle velocity simplifies into a motion solely along $\mathbf{B}$ such that
\begin{equation}
\label{eq:VitesseB}
\mathbf{v}_\pm =  \pm c \, \frac{( \mathbf{E} \cdot \mathbf{B} ) \, \mathbf{B}}{E_0 \, (E_0^2/c^2+B^2)} .
\end{equation}
This expression can be reduced to
\begin{equation}
\mathbf{v}_\pm = \pm c \, \textrm{sign}(B_0) \, \frac{\mathbf{B}}{B}
\end{equation}
by noting that in this weak electric field limit the magnitude of $\mathbf{B}$ is almost equal to the invariant $B_0$, namely $B^2 \approx B_0^2$. Particles are accelerated mostly by the electric component parallel to the magnetic field. The surfaces $\mathbf{E} \cdot \mathbf{B}=0$ are of particular interest because the velocity changes sign when the particle crosses this region. It is called a force-free surface and represents trapping regions for those particles \cite{ jackson_new_1976, finkbeiner_effects_1989, michel_electrodynamics_1999}.

The concept of magnetic field line in vacuum is misleading and specifying motion along a particular field line is not well defined in the general case. This requires some caution about the interpretation of the corotation speed $\mathbf{V}_{\rm rot}$. The way to follow the particle trajectory replaces this velocity by a special frame in which the electric field is parallel to the magnetic field leading to the Aristotelean dynamics discussed before. Indeed, the velocity $\beta_\parallel \, c$ required by the Lorentz transformation to get this condition is \cite{landau_physique_1989}
\begin{equation}
\frac{\boldsymbol{\beta}_\parallel}{1+\beta_\parallel^2} = \frac{c \, \mathbf{E} \wedge \mathbf{B}}{E^2 + c^2\, B^2}
\end{equation}
neglecting all other curvature, gradient and polarization drifts in the limit of vanishing Larmor radius which is correct in a super strong magnetic field. In that frame, where quantities are denoted by a prime, motion is along the common direction of $\mathbf{E}'$ and $\mathbf{B}'$. To get the useful solution, we write the frame velocity as
\begin{equation}
\label{eq:Vparallel}
\mathbf V_\parallel = \frac{\mathbf E \wedge \mathbf B}{E_0^2/c^2 + B^2} .
\end{equation}
The electric and magnetic fields in the frame moving at speed $\mathbf{V}_\parallel$ are found by a special-relativistic Lorentz boost of the electromagnetic field and gives
\begin{subequations}
	\begin{align}
	\mathbf E' & = \Gamma \, \frac{E_0}{E_0^2/c^2 + B^2} \, \left[ \frac{E_0}{c^2} \, \mathbf E + B_0 \, \mathbf B \right] \\
	\mathbf B' & = \Gamma \, \frac{B_0}{E_0^2/c^2 + B^2} \, \left[ B_0 \, \mathbf B + \frac{E_0}{c^2} \, \mathbf E \right] .
	\end{align}
\end{subequations}
Electric and magnetic fields are indeed collinear because $E_0 \, \mathbf B' = B_0 \, \mathbf E'$. In this frame, particles move along the common direction of $\mathbf{E}'$ and $\mathbf{B}'$. Thus the local tangent vector to the trajectory becomes $\mathbf{t}'_\parallel = \pm \mathbf{E}'/E' = \pm \mathbf{B}'/B'$, the sign being chosen such that particles flow outwards. Therefore we replace $\boldsymbol{\beta}$ by $\mathbf{V}_\parallel/c$ in eq.~(\ref{eq:AberrationLorentz}) to get a velocity field that should not be confused or seen as motion along field lines because this concept is usually ill defined for non-ideal plasmas when $\mathbf{E} \cdot \mathbf{B} \neq 0$. Our expression for the particle velocity resembles to the Aristotelian expression given by \cite{gruzinov_aristotelian_2013}. Our velocity prescription is however more general because we do not assume that particles travel exactly at the speed of light. The speed along the common $\mathbf{E}$ and $\mathbf{B}$ direction is unconstrained and fixed by the ``user'' contrary to Aristotelian electrodynamics.

If particles exactly move at the speed of light, in the comoving frame this velocity becomes $\mathbf v'= \pm \mathbf E'/E' = \pm \mathbf B'/B'$, the sign depending on the charge. Note also that the electromagnetic field strengthes are $E'=E_0$ and $B'=B_0$. Doing the Lorentz transformation to the observer frame, noting that $\mathbf V_\parallel$ and $\mathbf v'$ are orthogonal, this is nothing but Aristotelian electrodynamics. Our treatment is more general because we do not enforce the speed of light in this frame. The prescription for the velocity impacts the high-energy light-curves from pulsars. This has been shown in depth by \cite{petri_general-relativistic_2018}.

The parallel velocity~$\mathbf V_\parallel$ in eq.~(\ref{eq:Vparallel}) generalizes the electric drift approximation to field configurations with an electric field $E$ exceeding the magnetic field $c\,B$. There is no need to impose the condition $E<c\,B$ to respect the force-free condition. However, it reduces to force-free if $E_0=0$, meaning no radiation reaction and no dissipation meanwhile requiring $\mathbf{E} \cdot \mathbf{B} = 0$. 
In order to look for plasma filled magnetospheres, we have to resort to numerical simulations in the force-free regime or in a dissipative regime because of resistivity and/or radiation damping. In the next section, we show some new results for radiative pulsar magnetospheres to be compared with the standard force-free solution for oblique rotators.

\section{Numerical simulation of rotating magnetospheres}
\label{sec:NumericalSimulations}

Neutron stars cannot be surrounded by vacuum because particles are expelled from the surface and accelerated in the surrounding strong electromagnetic field. This is indirectly deduced from their broad band electromagnetic spectrum for which the Crab pulsar is an archetypal example \cite{moffett_multifrequency_1996}. Although an exact analytical solution for a rotating dipole in vacuum exists, known as Deutsch solution \cite{deutsch_electromagnetic_1955}, realistic magnetospheres require the presence of plasma producing charges and currents that retroact to the electromagnetic field. Because the problem is highly non linear, numerical simulations are compulsory. Two dimensional neutron star magnetospheres have been computed in the force-free regime two decades ago starting with the aligned FFE case \cite{contopoulos_axisymmetric_1999} and followed several years later by the general three dimensional oblique cases by \cite{spitkovsky_time-dependent_2006}. Since then, these results have been retrieved by several other authors using different numerical approaches like finite difference/finite volume methods \cite{komissarov_simulations_2006, timokhin_force-free_2006, kalapotharakos_extended_2012} or pseudo-spectral methods \cite{petri_pulsar_2012,cao_spectral_2016, cao_oblique_2016}. Even a combined spectral/discontinuous Galerkin method has been tried including general-relativistic effects for a monopole \cite{petri_general_2015} or a dipole \cite{petri_general-relativistic_2016}. Some extension to dissipative magnetospheres was undertaken by \cite{li_resistive_2012, cao_oblique_2016, kalapotharakos_toward_2012} assuming an ad hoc prescription for the dissipation. 

Here we show three models of pulsar magnetosphere for an oblique rotator with obliquity (angle between rotation axis and magnetic axis) $\chi = \{0\degree, 30\degree, 60\degree, 90\degree\}$, namely the vacuum, the force-free and the radiative cases. Simulations are performed in the observer inertial frame but using the corotating coordinate system leading to Maxwell equations written as eq.~(\ref{eq:MaxwellCorotation}). This particular frame ensures that the solution relaxes to a time independent solution where the current sheet remains at a fixed position in space in order to ease its location for subsequent purposes. In other words, the time derivatives in eq.~(\ref{eq:MaxwellCorotation}) must vanish when the solution becomes stationary.

The three models correspond to three prescriptions for the electric current density~$\mathbf j$. In vacuum, for the Deutsch solution it is obviously $\mathbf j = \mathbf 0$. This is our reference solution for checking our algorithm and accuracy of the computed solution. Simulations are performed using our pseudo-spectral Maxwell solver explained in depth in \cite{petri_pulsar_2012}. Before discussing our new results, we remind the essential features of our pseudo-spectral code in the following paragraph.

\subsection{Numerical schemes}

Spectral and pseudo-spectral numerical schemes convert a system of partial differential equations (PDE) into a larger system of ordinary differential equations (ODE) much easier to integrate numerically with standard ODE integration techniques like the explicit Runge-Kutta and Adams-Bashforth schemes. See \cite{canuto_spectral_2006} for a detailed review on these techniques. We emphasize that spectral methods do not approximate the equations of the problem but the solution itself. Therefore the numerical problem exactly reflects the mathematical problem with the same boundary conditions which need to be properly imposed without any under or over-determinacy. Note that finite volume/finite difference codes are prone to large (with respect to spectral codes) diffusion/dissipation and are therefore able to damp boundary conditions that are not exactly identical to the mathematical problem making it analytically an ill-posed problem (mathematically speaking not from a numerical point of view). Spectral methods are primarily dealing with expansion coefficients of the unknown quantities not their value themself at the grid points. This expansion possesses the great advantage of removing singularities of differential operators like the gradient, the divergence and the curl in spherical coordinates along the polar axis. We use this flexibility to solve Maxwell equations in polar spherical coordinates $(r,\theta,\varphi)$ with no special care about the polar axis. Boundary conditions on the stellar surface can thus be properly and exactly imposed as required by the original mathematical problem.

Specifically, the components of the electromagnetic field are expanded onto a real Fourier-Legendre-Chebyshev basis. The azimuthal dependence is expanded into a standard Fourier series in $\cos(m\,\varphi)$ and $\sin(m\,\varphi)$ whereas the latitude is expanded into Legendre functions $P_\ell^m(\theta)$ where $\ell$ and $m$ are integers related to the spherical harmonics $Y_{\ell,m}(\theta, \varphi)$ \cite{arfken_mathematical_2005}. The radial part is expanded into Chebyshev polynomials $T_n(x(r))$ where $r \in [R_1, R_2]$ is mapped into the normalized range~$x\in[-1,1]$ by a linear transformation. The straightforward implementation of this mapping accumulates the discrete grid found from the Chebyshev-Gauss-Lobatto points unevenly near the boundary points where the resolution becomes prohibitively high. The constrain on the time step is therefore to severe. In order to distribute more evenly the grid points, we use the Kozloff/Tal-Ezer mapping \cite{kosloff_modified_1993}. See also \cite{parfrey_introducing_2012} for a similar implementation of this technique for axisymmetric neutron star magnetospheres. Derivatives are computed in the Fourier-Legendre-Chebyshev space by simple algebraic operations, instead of pure function derivatives, and then transformed back to real space on grid points. The outer boundary conditions are outgoing waves with a sponge layer absorbing spurious reflections. The inner boundary conditions enforce the tangential part of the electric field and the normal component of the magnetic field at the stellar surface. To keep a mathematically well-posed problem, we employ the characteristic compatibility method described in \cite{canuto_spectral_2006}. Time integration is performed via a standard third order Runge-Kutta scheme.

Spectral methods are known to converge to the exact solution faster than finite difference or finite volume schemes for sufficiently smooth problems without discontinuities. They require less resolution for the same accuracy \cite{boyd_chebyshev_2001}. Because spectral methods rely on Fourier-like series expansions, they are also sensitive to the Gibbs phenomenon \cite{arfken_mathematical_2005}, spoiling the solution with overshoot possibly leading to unphysical quantities like negative densities or pressures. In our strong electromagnetic field limit, however, no positivity constrain is required for the unknown field. However, in order to stabilize the algorithm, tending to put more and more energy into small scales because of the Gibbs effects, we need to filter the highest frequencies by applying for instance an exponential filter damping the highest order coefficients in the Fourier-Legendre-Chebyshev expansion. Eventually, we check a posteriori that the simulation has converged to the desired solution to good accuracy by performing a resolution analysis, meaning that increasing by a factor two the grid resolution in each direction, the solution does not significantly changes. We found that for the simulations shown below, a resolution $N_r\times N_\theta \times N_\varphi = 257 \times 32 \times 64$ already gave reasonable results. We checked on a few cases that increasing by a factor 2 the resolution in all directions did not change the results (but drastically increased the computational time on a single core). Consequently, we adopted a resolution of $N_r\times N_\theta \times N_\varphi = 257 \times 32 \times 64$ for accurate and converged results.
In the special case of a aligned rotator, the Gibbs phenomenon is strongest. We had to resort to higher resolution of $N_r\times N_\theta \times N_\varphi = 513 \times 64 \times 1$ for accurate and converged results.

\subsection{Force-free magnetospheres}

In the force-free regime where $\mathbf{E} \cdot \mathbf{B}=0$ and $E<c\,B$, the electric current density is uniquely defined by \cite{gruzinov_stability_1999}
\begin{equation}
\label{eq:J_Ideal}
\mathbf j = \rho_{\rm e} \, \frac{\mathbf{E}\wedge \mathbf{B}}{B^2} + \frac{\mathbf{B} \cdot \rot \mathbf{B} / \mu_0 - \varepsilon_0 \, \mathbf{E} \cdot \rot \mathbf{E}}{B^2} \, \mathbf{B} .
\end{equation}
This current is decomposed into an electric drift part, first term on the right hand side, depending only on the total electric charge density~$\rho_{\rm e}$, and a part along the magnetic field that is not constrained but deduced a posteriori from the simulation output. Because all particles drift with the same velocity, contribution to the drift part of the electric current arises solely from the non-neutrality of the plasma, meaning $\rho_{\rm e}\neq0$.

The electric drift speed must remains strictly less than the speed of light. If the condition $E<c\,B$ is violates, force-free breaks down and the plasma becomes dissipative. In force-free simulations however, we enforce by hand the condition $E<c\,B$ everywhere in space in order to stay in the sub-relativistic drift speed limit.

\begin{figure}[H]
	\centering
	\includegraphics[width=0.9\linewidth]{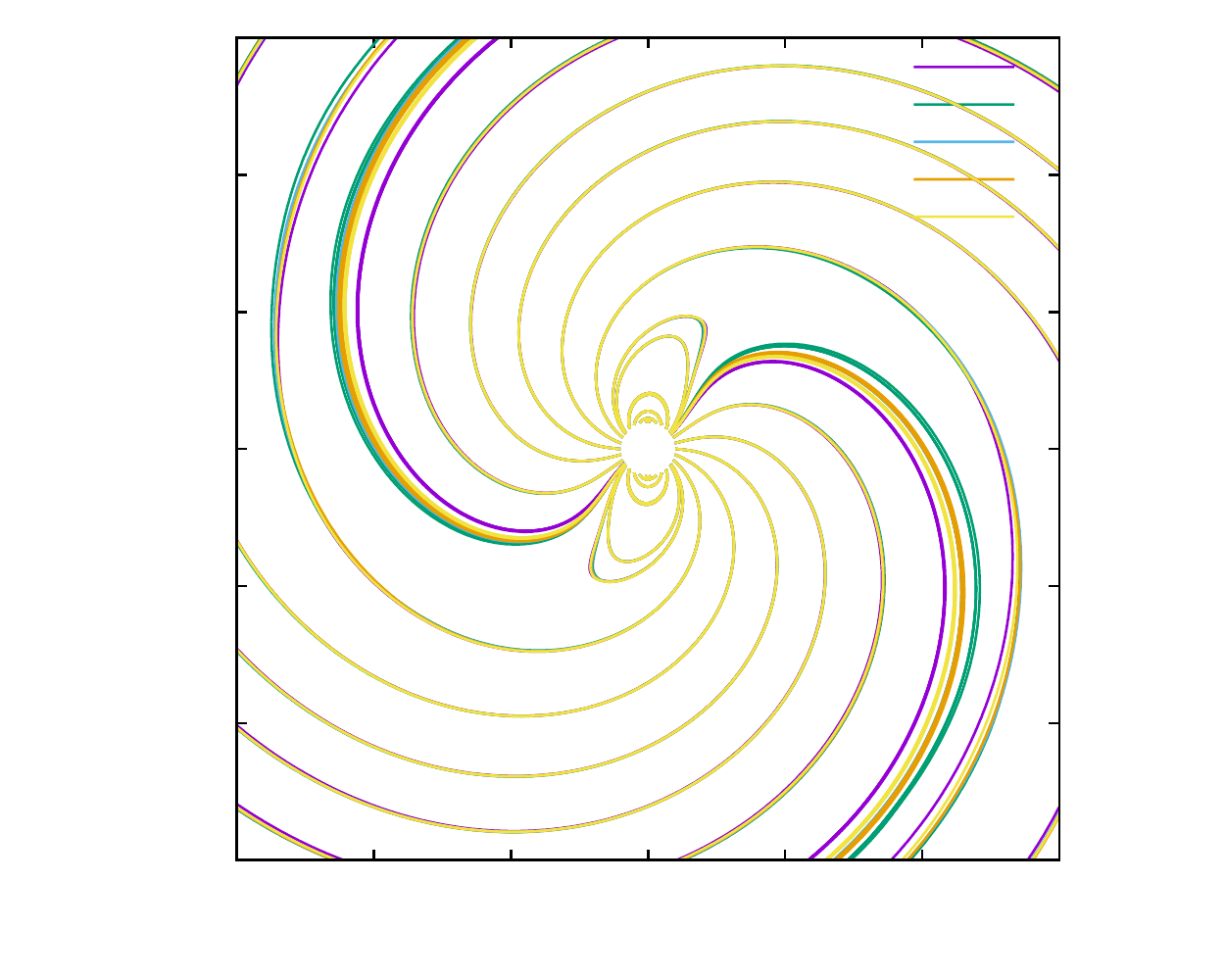}
	\caption{Magnetic field lines for orthogonal ($\chi=90\degree$) force-free (FFE) and radiative magnetospheres for different values of pair multiplicity~$\kappa$.\label{fig:lignes_ffe}}
\end{figure}

An example of magnetic field lines in the equatorial plane for an orthogonal rotator with $\chi=90\degree$ is shown in figure~\ref{fig:lignes_ffe} with $R/\rlight=0.2$. The force-free regime tries to put the field lines out of the current sheet which becomes singular, but due to numerical resistivity, field lines also tend to close by crossing this singular surface. 

\subsection{Radiative magnetospheres}

For the radiative magnetosphere, we introduce a additional free parameter represented by the pair multiplicity factor~$\kappa$ such that the electric current derived from the Aristotelian electrodynamics becomes
\begin{equation}
\mathbf j = \rho_{\rm e} \, \frac{\mathbf E \wedge \mathbf B}{E_0^2/c^2 + B^2} + |\rho_{\rm e}| \, (1+2\,\kappa) \, \frac{E_0 \, \mathbf E/c^2 + B_0 \, \mathbf B}{E_0^2/c^2 + B^2} .
\end{equation}
It is decomposed into a $\mathbf E \wedge \mathbf B$ drift similar to force-free but without the additional constraint $E<c\,B$ and a part along $\mathbf E$ and $\mathbf B$ which reduces in the drift frame to a motion along the common direction of $\mathbf E'$ and $\mathbf B'$.

Fig.~\ref{fig:lignes_ffe} shows some field lines in the equatorial plane for the orthogonal rotator with $\chi=90\degree$ in the radiative regime with pair multiplicity~$\kappa=\{0,1,2,5\}$. In the most dissipative case corresponding to $\kappa=0$, field lines cross the current sheet at smaller distances compared to less dissipative cases with $\kappa=2$ or $\kappa=5$.

\begin{figure}[H]
	\centering
	\includegraphics[width=0.9\linewidth]{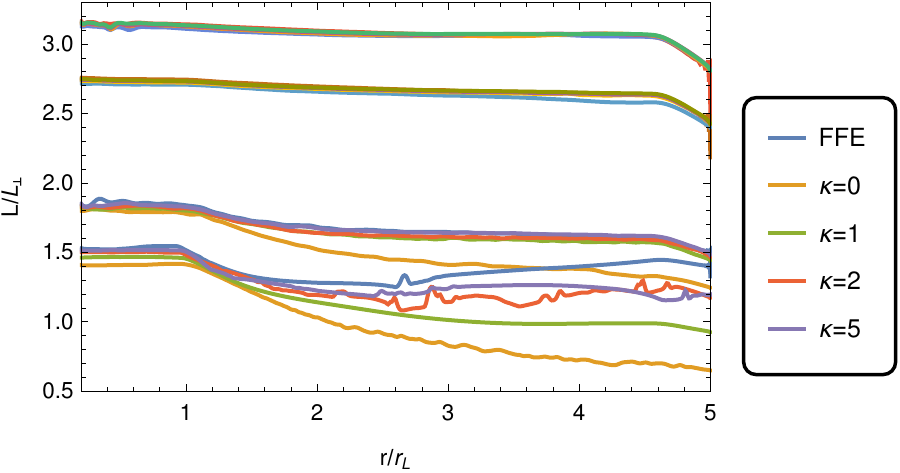}
	\caption{The radial dependence of the Poynting flux for an oblique rotator in force-free and radiative regimes. \label{fig:luminosite_radiale}}
\end{figure}

Fig.~\ref{fig:luminosite_radiale} shows the associated radial dependence of the Poynting flux for force-free and radiative cases with $\kappa=\{0,1,2,5\}$. The radiative magnetosphere dissipates a small fraction of the Poynting flux into particle acceleration and radiation, most efficiently when $\kappa=0$, corresponding to a charge separated plasma. Increasing the pair multiplicity factor~$\kappa$ to higher values shifts the radiative model towards the force-free limit. In the aligned case, the decrease in Poynting flux is abrupt right at the light-cylinder. It is most prominent for $\kappa=0$. However, due to the intrinsic dissipation of our algorithm, even in the FFE case there some Poynting flux dissipation is observed. This is due to the infinitely thin current sheet with discontinuous toroidal magnetic field that is smeared by our spectral methods (Gibbs phenomenon). The situation improves for oblique cases as the displacement current take over some fraction of the electric current within the sheet.

\begin{figure}[H]
	\centering
	\includegraphics[width=0.9\linewidth]{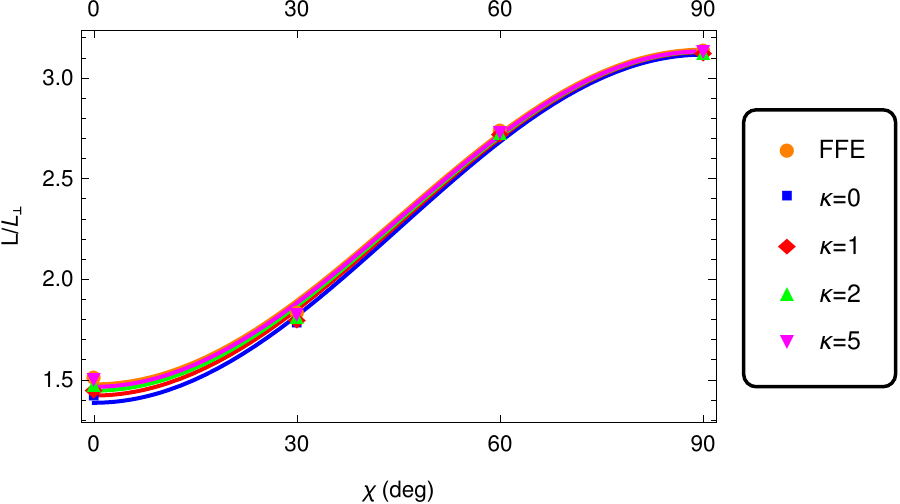}
	\caption{The Poynting flux crossing the light-cylinder for oblique rotators in force-free and radiative regimes corresponding to fig.~\ref{fig:luminosite_radiale}.\label{fig:luminosite_oblique}}
\end{figure}

In fig.~\ref{fig:luminosite_oblique} we show the Poynting flux crossing the light-cylinder for force-free and radiative cases with $\kappa=\{0,1,2,5\}$ and depending on the inclination angle~$\chi$. All cases can be fitted with a single formal expression summarized as
\begin{equation}\label{eq:fit}
 L = L_\perp \, ( a + b \, \sin^2 \chi )
\end{equation}
with different coefficients depending on the regime considered. The fitted values extracted from the numerical simulations are listed in Table \ref{tab:Fit_Spindown}. The most dissipative case $\kappa=0$ slightly decreases the Poynting flux for the aligned rotator already inside the light-cylinder. The decrease is accurately quantified by the fitting parameter~$a$. The FFE normalized Poynting flux is 1.42 whereas for the radiative $\kappa=0$ case it is 1.36. The fitting parameter~$b$ seems less dependent to the regime considered. The aligned rotator also shows the most prominent gradual decrease in the Poynting flux with respect to distance. Dissipation starts at the light cylinder but goes on at several light cylinder radii. For oblique rotators, the slope of this radial decrease slowly diminishes, becoming negligible for the orthogonal rotator.
\begin{table}
	\centering
\begin{tabular}{|ccc|}
\hline 
Regime &	a & b \\
\hline 	\hline 
FFE & 1.42 & 1.73 \\
$\kappa=0$ & 1.36 & 1.75 \\
$\kappa=1$ & 1.39 & 1.76 \\
$\kappa=2$ & 1.40 & 1.74 \\
$\kappa=5$ & 1.42 & 1.73 \\
\hline 
\end{tabular} 
\caption{Fitting coefficients $a$ and $b$ for the spin-down luminosity as fitted in eq.~(\ref{eq:fit}). \label{tab:Fit_Spindown}}
\end{table}

In all regimes, the electromagnetic fluxes are very similar while inside the light-cylinder. The discrepancies occur outside the light-cylinder, in regions where the electric field is dominant and not fully screened by the plasma because of the too low pair multiplicity. A corotative ideal and dissipationlessness magnetosphere inside the star is therefore a good approximation, whereas outside, efficient dissipation sets in right at the light-cylinder, around the current sheet.

\begin{figure}[H]
	\centering
	\includegraphics[width=0.9\linewidth]{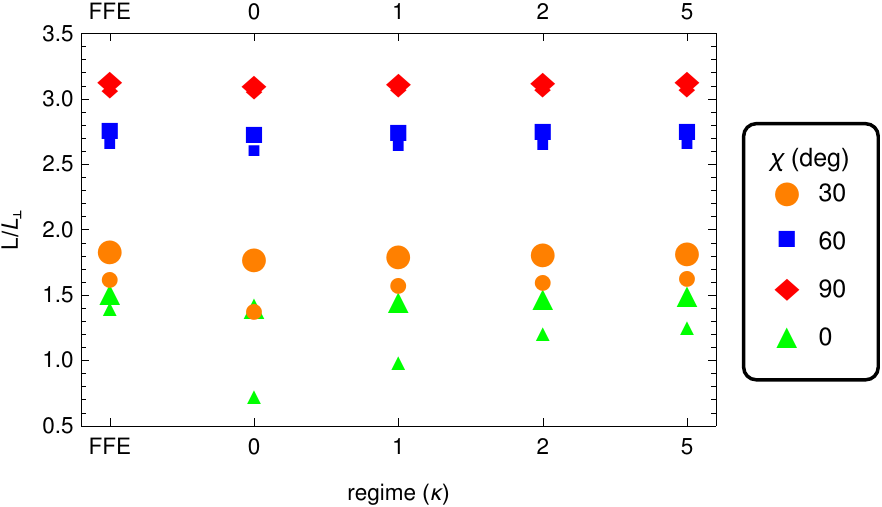}
	\caption{The Poynting flux crossing the light-cylinder (larger markers) and crossing a sphere of radius $4\,\rlight$ (smaller markers) for oblique rotators in force-free and radiative regimes.\label{fig:luminosite_regime}}
\end{figure}

Fig.~\ref{fig:luminosite_regime} shows a summary of the Poynting flux crossing the light-cylinder (larger markers) and crossing a sphere of radius $4\,\rlight$ (smaller markers) for oblique rotators in force-free and radiative regimes. The dissipation going on at large distances is most visible for the aligned rotator with green triangles.

\begin{figure}[H]
	\centering
	\begin{tabular}{cc}
	\includegraphics[width=0.5\linewidth]{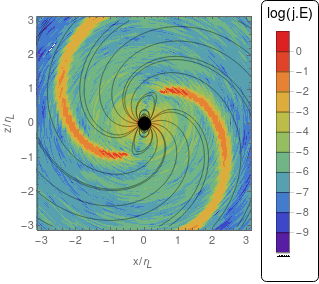} &
	\includegraphics[width=0.5\linewidth]{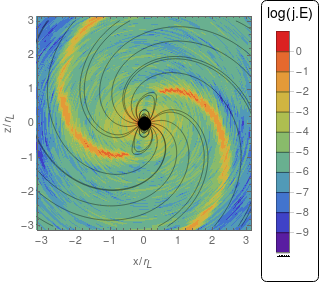} \\
	\includegraphics[width=0.5\linewidth]{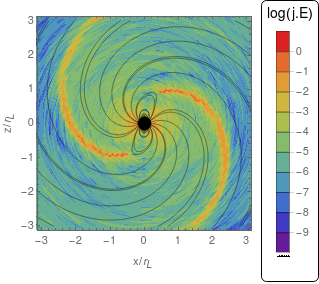} &
	\includegraphics[width=0.5\linewidth]{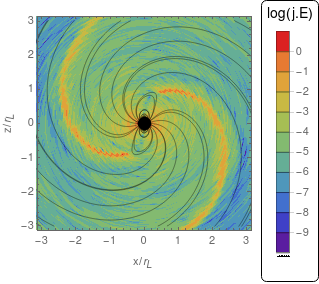}
	\end{tabular}
	\caption[Dissipation]{Dissipation in the equatorial plane of an orthogonal rotator for $\kappa=\{0,1,2,5\}$ (from left to right, top to bottom).}
	\label{fig:dissipationr0}
\end{figure}

Some fraction of the electromagnetic flux goes into particle acceleration and radiation. Quantitatively, this dissipation of the electromagnetic energy is computed as a work done on the plasma such that
\begin{equation}
\label{eq:jscalaireE}
\mathbf{j} \cdot \mathbf{E} = |\rho_e| \, ( 1 + 2 \, \kappa ) \, c \, E_0 \geq 0 .
\end{equation}
This dissipation rate, for $\kappa=\{0,1,2,5\}$, is shown in Fig.~\ref{fig:dissipationr0} on a log scale. It shows the location of largest dissipation for an orthogonal rotator according to the dissipation rate controlled by $\kappa$. Poynting flux goes into particle acceleration and radiation mainly outside the light-cylinder along the current sheet starting from the Y-point. We expect therefore gamma-rays to be produced along this sheet, emitting pulses at the neutron star rotation frequency. Such models have already been put forward and known as the striped wind. See for instance \cite{lyubarskii_model_1996} for the production of high-energy emission and \cite{kirk_pulsed_2002} for demonstrating the pulsation.
We conclude that energy conversion occurs mainly around the current sheet. Within the light-cylinder, the electric field is always less than the magnetic field $E<c\,B$. Therefore the force-free condition can be maintained without resorting to artificial damping of $\mathbf{E}$. However, dissipation sets in right at the light-cylinder, where magnetic field lines start to cross the light-cylinder. The dissipation region follows a spiral pattern with decreasing amplitude with distance from the star. These new simulations offer for the first time a fully self-consistent description of a dissipative and radiative magnetosphere, where feedback between plasma flow, particle radiation and electromagnetic field is included. Note that emission occurs only along the current sheet outside the light-cylinder. This conclusion supports the idea of the striped wind model introduced by \cite{michel_magnetic_1994} and by \cite{coroniti_magnetically_1990}. It also explains pulsed high-energy emission from gamma-ray pulsars as demonstrated by \cite{kirk_pulsed_2002}, \cite{petri_polarization_2005} and \cite{petri_unified_2011}.

Contrary to the vacuum case, by construction, particles cannot move faster than the speed of light, even if the corotating velocity eq.~(\ref{eq:AberrationCorot}) is used. This is because the magnetic field is now sufficiently bent to counterbalance the effect of adding the corotation velocity given by eq.~(\ref{eq:VitesseCorotation}). Note also that radiative magnetospheres presented in our study do not tend to the vacuum solution when the pair multiplicity vanishes $\kappa=0$ because inside the light-cylinder we enforce force-free conditions by construction.

Dissipative losses in the current sheet, also called striped wind, have also been proposed by other authors. For instance \cite{contopoulos_new_2014} found a new standard solution for the aligned rotator, free of separatrix current layer within the light-cylinder. Dissipation occurs only in the equatorial current sheet where acceleration and radiation of particle is allowed. They found an increase of 23\% of the spindown with respect to $L_\perp$, 40\% of which goes into the current sheet dissipation. In our solutions, we found a spindown increase from 36\% to 42\% depending on the pair multiplicity, see table~\ref{tab:Fit_Spindown}. The crux of the matter is the microphysical description of this current sheet that conditions the whole magnetospheric solution. In order to prescribe the electric current in this sheet, \cite{contopoulos_new_2014} assumed a null-like current everywhere, a prescription which is questionable. Moreover 60\% of the magnetic flux crossing the light-cylinder opens up to infinity. It is not clear how this percentage is controlled by the solution. A better solution would get all magnetic flux dissipated sooner or later in the equatorial current sheet. \cite{philippov_ab_2014} used another approach, performing Particle In Cell (PIC) simulations of pulsar magnetospheres. Here the sensitive parameter is the unconstrained pair injection rate, from the surface or from the whole magnetosphere. The stationary solution crucially depends on this injection mechanisms, going from an electrosphere to an almost force-free magnetosphere. They found that less than 15\% of the Poynting flux is dissipated within $2\,\rlight$. It is not clear how much additional decrease is expected if the solution would have been computed to larger distances. A partial answer is given in \cite{cerutti_particle_2015} where the dissipation is as high as 35\% at $5\,\rlight$. Comparing both models is difficult because they are not performed with the same set up. The most critical variable being the pair multiplicity which is not fixed by the user and not easily controlled. We showed that $\kappa$ strongly affects the asymptotic large distance dissipation in the axisymmetric case. These different approaches can only be reconciled in light of the pair content within the magnetosphere.

To summarize, all results performed with different numerical codes and different assumptions demonstrated that the magnetosphere relaxes automatically to a state where corotation with the star is enforce by the electric current prescription. However, while this picture is simple and easily understood, nothing forbids solutions with differentially rotating plasmas. Such solutions are discussed in the next section.

\section{Differentially rotating magnetospheres}
\label{sec:DifferentialMagnetospheres}

The neutron star magnetosphere is often described as perfectly corotating with the star, dragged by the electromagnetic field to enforce strict corotation as in the simulations performed in the previous section. However, it is well known from Ferraro isorotation law \cite{ferraro_non-uniform_1937} that to keep corotation, the plasma must be connected magnetically to the star everywhere in the magnetosphere. If some vacuum gaps exist between the surface and the magnetospheric plasma, corotation becomes impossible. The plasma around the equatorial plane will start to rotate differentially, leading to a much complexer variety of physical processes like non neutral plasma instabilities \cite{davidson_introduction_1990} and efficient particle diffusion across magnetic field lines \cite{dubin_collisional_1998, dubin_trapped_1999}.

A perfectly corotating magnetospheric plasma, as simple as it could be, does not represent a realistic pulsar magnetosphere. Differential rotation or lagging of particle motion is permitted when vacuum gaps are allowed. Indeed, in the electrospheric solution found by \cite{krause-polstorff_electrosphere_1985} and detailed by \cite{petri_global_2002} for an aligned rotator, the domes are corotating because magnetically connected to the star but the equatorial disc over-rotates with respect to the star. This differential rotation is induced by a charge density given for an aligned rotator by
\begin{equation}\label{key}
\rho = - \varepsilon_0 \, (2 \, \mathbf{\Omega} \cdot \mathbf{B} + r^2 \, B^2 \, \Omega'(a)) 
\end{equation}
where $a$ is the magnetic flux function. The quantity $\rho_{\rm gj} = - 2 \, \varepsilon_0 \, \mathbf{\Omega} \cdot \mathbf{B}$ is usually referred as the Goldreich-Julian charge density \cite{goldreich_pulsar_1969}. The charge density $\rho$ in absolute value is much higher than the one required for corotation \cite{hones_electric_1965} because $\Omega'(a)>0$. Therefore, the location of the light cylinder shifts nearer towards the surface. Actually it is no more a cylinder but an azimuthally symmetric surface. In the general case, it is more suitable to speak about a light-surface rather than about a light-cylinder.

This new structure has profound consequence on the secular evolution of the magnetosphere. Indeed, \cite{petri_diocotron_2002} showed that the equatorial disc is unstable with respect to the diocotron instability. A quasilinear theory has been developed by \cite{petri_cross-field_2003} demonstrating the possibility to transport charges across magnetic field lines. This is of paramount importance for the magnetosphere. Later \cite{petri_non-linear_2009} investigated through electrostatic PIC simulations the fully non-linear evolution of the diocotron instability. He found that particles are transported radially outwards in the equatorial plane when the system is feed with fresh electron/positron pairs from the innermost part of the magnetosphere (produced for instance by magnetic photon absorption or photon-photon collisions). However, \cite{petri_relativistic_2007} proved that the relativistic rotation tends to stabilize the diocotron instability. Moreover particle inertia becomes significant close to the light-cylinder, but the instability still survives, switching to the magnetron case \cite{petri_magnetron_2008}.

The presence of such instabilities destroys the picture of a stationary and corotating magnetosphere. By nature the pulsar electrodynamics is non stationary as can be witnessed from the highly erratic emission feature of single radio pulse profiles \cite{manchester_radio_2009}. This is inherent to the relativistic plasma flow and to the pair creation process occurring close to the stellar surface and/or close to the light-cylinder. This remark leads us to the last topic concerning radiative signatures within the magnetosphere.

\section{Radiation from the magnetosphere}
\label{sec:Radiation}

Pulsars show a broadband emission from radio through optical up to X-rays and gamma-rays. Photons are produced within the magnetosphere and wind. They must indirectly carry information about their production site, therefore showing an imprint of relativistic rotation if produced close to the light-cylinder. 

The location of high-energy emission from gamma-ray pulsars is poorly constrained by observations. Several competing models interpret the measurements within the magnetosphere or wind, with either curvature, synchrotron or inverse Compton radiation. So far, there is little hope to get a clear insight about the magnetosphere from these observations. Much more interesting in our view is the wealth of data in radio pulse profiles and their associated polarisation feature. Radio emission height in normal radio pulsars with slow periods, larger than 100~ms, is well constrained to lie at altitudes around several hundreds of kilometres \cite{mitra_nature_2017}. This is deduced from the shift in polarization angle traverse with respect to the middle of the pulse profile. This shift~$\Delta \phi$ is explained in the framework of aberration/retardation effects of photons propagating in the magnetosphere \cite{blaskiewicz_relativistic_1991} and amounts to
\begin{equation}\label{eq:ARformula}
\Delta \phi \approx 4\,r/\rlight .
\end{equation}
This shift cannot be explained by emitting a photon along a field line in the corotating frame and then using aberration formulas because the pulse profile would be subject to the same shift in phase as the polarization angle traverse and therefore cancelling the possible time delay between the middle of the pulse profile and the steepest gradient in the polarization angle. The only way to correctly catch this shift, which is a well defined fact seen in many observations of radio pulsars, is through photons emitted along field lines dragged into corotation as seen in the observer inertial frame. This corotation velocity leads naturally to a time lag between the middle of the pulse profile and the polarization angle inflexion point. Nevertheless, the magnetic field topology has to adjust to compensate for the corotation velocity in order to keep the particle velocity less than the speed of light. This is not always possible outside the light-cylinder as already explained in paragraph~\ref{subsec:velocity}. The approximate estimate given in eq.(\ref{eq:ARformula}) has been checked for off-centred and rotating dipoles by \cite{petri_joined_2019}. It is therefore a very robust result, sharply constraining the radio emission heights.

Consequently, the view presented in paragraph~\ref{subsec:frame} must be rejected because it cannot reproduce aberration/retardation effects in pulsar radio polarization observations. However, the particle motion described in paragraph~\ref{subsec:velocity} seems more appropriate. But the best choice in our view is represented in paragraph~\ref{subsec:aristote} where trajectories are computed according to the full electromagnetic field taking into account radiative effects. The latter option gives the simplest plausible scenario where radiation reaction acts efficiently and self-consistently backwards onto the particle motion and onto the electromagnetic field.

%
%


\section{Discussion}
\label{sec:Discussion}

Pulsar magnetospheres have been extensively computed for stellar centred dipolar fields in special relativity. However, there are increasing evidences for off-centred or even multipolar components anchored in the neutron star crust. Indeed, joined modelling of pulsed radio emission and thermal X-rays from the polar cap hot spots requires decentred dipoles \cite{petri_joint_2020}. Moreover detailed investigations of X-ray light curves of millisecond pulsars also favours off-centred and quadrupolar components according to recent observations from NICER \cite{bilous_nicer_2019}. Nevertheless, as shown by \cite{petri_illusion_2019}, we do not expect drastic changes in the spin-down luminosity and magnetic field structure outside the light-cylinder for slowly rotating pulsars with period $P>10$~ms. In the case of radiative magnetospheres, we expect a similar trend because radiation occurs outside the light-cylinder where the multipolar components have sufficiently decreased to become negligible. Indeed, a dipole magnetic field decreases like $1/r^3$ whereas a multipole of order~$\ell$ decreases like $1/r^{\ell+2}$. Therefore a multipole of strength $B_{\rm m}$ at the surface contributes only a ratio $(B_{\rm m} / B_{\rm d} ) \, (R/\rlight)^\ell$ compared to a dipole of strength $B_{\rm d}$ at the surface. We can draw the same conclusions for general-relativistic radiative magnetospheres. Indeed, for force-free magnetospheres, \cite{petri_general-relativistic_2015} already showed that qualitatively the picture does not vary and that the spindown luminosity scales like the magnetic field strength at the light-cylinder. Extrapolating to the radiative case, general-relativistic effects remain very weak outside the light-cylinder for any pulsar, millisecond or second and the overall picture discussed above remains valid.

\section{Conclusions}
\label{sec:Conclusions}

Rotation in neutron stars plays a central role to sustain their electromagnetic activity of particle acceleration, pair creation and the subsequent broadband radiation from radio wavelengths to very high energies. A corotating magnetosphere is often used as a good approximation to describe its electrodynamics. However, due to their fast rotation, relativistic speeds are reached already close to the stellar surface, around the light-cylinder. This hypothetical surface separates the inner magnetosphere from the wind. We showed that the transition zone between the quasi-static magnetosphere and the wave zone in the wind is difficult to treat satisfactorily and smoothly because of the absence of a physical frame rotating with the star outside this light cylinder. Several prescriptions where exposed, starting from different assumptions. We also showed that pulsar radio observations give some hint about promising paths to follow particle trajectories and their emission.

Clearly, a deeper and better understanding of the neutron star electrodynamics is required to faithfully explain the wealth of data about their radiation. Neutron stars are one of the only macroscopic objects that produce relativistic rotation on a length scale of the order of Earth radius about several hundreds to several thousands of kilometres. Investigating jointly theoretical models and observational facts will reveal relativistic rotation effects in astrophysics in an unusual way, hoping to solve half a century mystery bout their functioning.


\vspace{6pt} 




\funding{This research was funded by CEFIPRA grant number IFC/F5904-B/2018. We would like to acknowledge the High Performance Computing center of the University of Strasbourg for supporting this work by providing scientific support and access to computing resources. Part of the computing resources were funded by the Equipex Equip@Meso project (Programme Investissements d'Avenir) and the CPER Alsacalcul/Big Data. We also thank the International Space Science Institute, Berne, Switzerland for their hospitality and for providing financial support during the meeting led by I. Contopoulos \& D. Kazanas that helped to improve the present work.}

\reftitle{References}





\end{document}